\documentclass[3p,twocolumn]{elsarticle}

\begin{document}

\begin{frontmatter}

\title{$^{11}$Be($\beta$p), a quasi-free neutron decay?}

\author[aar]{K. Riisager\corref{cor1}}
\ead{kvr@phys.au.dk}
\author[vie,smi]{O. Forstner}
\author[cern,iem]{M.J.G. Borge}
\author[iem]{J.A. Briz} 
\author[iem]{M. Carmona-Gallardo}
\author[com]{L.M. Fraile} 
\author[aar]{H.O.U. Fynbo} 
\author[ce2]{T. Giles}
\author[iem,ce2]{A. Gottberg} 
\author[got]{A. Heinz}
\author[aar]{J.G. Johansen\fnref{fn1}} 
\author[got]{B. Jonson} 
\author[cern]{J. Kurcewicz} 
\author[aar]{M.V. Lund}
\author[got]{T. Nilsson} 
\author[got]{G. Nyman} 
\author[cern]{E. Rapisarda}
\author[vie]{P. Steier}
\author[iem]{O. Tengblad} 
\author[got]{R. Thies} 
\author[vie]{S.R. Winkler}

\cortext[cor1]{Corresponding author}
\fntext[fn1]{Present address: Institut f\"{u}r Kernphysik, Technische Universit\"{a}t
 Darmstadt, D--64289 Darmstadt, Germany}

\address[aar]{Department of Physics and Astronomy, Aarhus University,
 DK--8000, Aarhus C, Denmark}

\address[vie]{Faculty of Physics, University of Vienna, W\"{a}hringer Strasse
  17, A--1090 Wien, Austria}

\address[smi]{Stefan-Meyer-Institut f\"{u}r subatomare Physik, Austrian Academy
  of Sciences, A--1090 Wien, Austria}

\address[cern]{ISOLDE, PH Department, CERN, CH--1211 Geneve 23, Switzerland  }

\address[iem]{Instituto de Estructura de la Materia, CSIC, E-28006 Madrid, Spain}

\address[com]{Grupo de F\'{\i}sica Nuclear, Universidad Complutense de Madrid,
CEI Moncloa, E-28040 Madrid, Spain}

\address[ce2]{EN Department, CERN, CH--1211 Geneve 23, Switzerland  }

\address[got]{Fundamental Fysik, Chalmers Tekniska H\"{o}gskola, SE--41296
  G\"{o}teborg, Sweden}

\begin{abstract}
  We have observed $\beta^-$-delayed proton emission from the neutron-rich
  nucleus $^{11}$Be by analysing a sample collected at the ISOLDE
  facility at CERN with accelerator mass spectrometry (AMS). With a
  branching ratio of $(8.4 \pm 0.6)\cdot 10^{-6}$ the strength of this
  decay mode, as measured by the $B_{GT}$-value, is unexpectedly high.
  The result is discussed within a simple single-particle model and
  could be interpreted as a quasi-free decay of the $^{11}$Be halo
  neutron into a single-proton state.
\end{abstract}


\end{frontmatter}


\section{Introduction}

Beta-minus decay and proton emission take a nucleus in almost opposite
directions on a nuclear chart, so $\beta^-$-delayed proton emission
(where beta decay feeds excited states that subsequently emit a
proton) is forbidden in all but a few nuclei where it is heavily
suppressed as the available energy is \cite{Jon01} $Q_{\beta p} = 782
\,\mathrm{keV} - S_n$ where $S_n$ is the neutron separation energy of
the nucleus. We describe here an experiment to detect this decay mode
from the one-neutron halo nucleus $^{11}$Be that is believed to be the
most favourable case \cite{Bay11,Bor13} due to the single-particle
behaviour of halo nuclei \cite{Jen04,Tan13,Rii13} that may favour this
decay mode and due also to the relatively long halflife that is caused
by the normal beta-decay of $^{11}$Be being hindered since a level
inversion gives it a $1/2^+$ ground state rather than a $1/2^-$.


Beta-delayed particle emission is in general a prominent decay mode for
nuclei close to the dripline, see \cite{Pfu12,Bla08} for recent
reviews. The energetically open channels for $^{11}$Be are
$\beta\alpha$, $\beta$t, $\beta$p and $\beta$n with corresponding
$Q$-values of \cite{mas12} $2845.2 \pm 0.2$ keV, $285.7 \pm 0.2$ keV,
$280.7 \pm 0.3$ keV and $55.1 \pm 0.5$ keV. The low decay energy implies
that the branching ratio for beta-delayed proton emission is low, typical estimates are
slightly above $10^{-8}$ \cite{Bor13}. 
To detect the process experimentally, it is therefore essential to keep contaminants
at a very low level.

The $\beta$p decay mode may be expected
preferentially in one-neutron halo nuclei, partly due to the
requirement of low neutron separation energy, partly due to the more
pronounced single-particle behaviour of halo nuclei. Two-neutron halo
nuclei are in a similar way candidates for beta-delayed deuteron
emission, which has so far been observed only in the nuclei
$^6$He and $^{11}$Li \cite{Pfu12,Nil00}. For $^{11}$Li the decay has a
branching ratio of order $10^{-4}$, the low value again caused by a
small energy window, whereas cancellation effects reduces the
branching ratio for $^6$He down to the $10^{-6}$ level. It may be more
useful to consider the standard
measure for the strength of a decay, the reduced matrix element
squared $B_{GT}$, that is found from the relation \cite{Pfu12}
\begin{equation}   \label{eq:ft}
  ft = \frac{K}{g_V^2 B_F + g_A^2 B_{GT}}
\end{equation}
where $f$ is the beta-decay phase space, $K/g_V^2= 6144.2 \pm 1.6$ s
and $g_A/g_V=-1.2694 \pm 0.0028$.  Converting the observed spectra for
beta-delayed deuteron emission from the two-neutron halo nuclei $^6$He
\cite{Ant02} and $^{11}$Li \cite{Raa08} gives total $B_{GT}$ values
within the observed energy range of about 0.0016 and 0.75. (Note,
however, that the $^6$He decay to the $^6$Li ground state has been
described as an effective di-neutron to deuteron decay, it is a
super-allowed transition with a $B_{GT}$ of 4.7. This may be a
reflection of a general trend for super-allowed decays to occur in
very neutron-rich nuclei \cite{Bor91}.) For comparison, the sum of
$B_{GT}$ for all currently known transitions in the $^{11}$Be decay is
0.27.

\section{The experiment}

\subsection{General remarks}
The radioactive $^{11}$Be nuclei were produced at the ISOLDE facility
at CERN.  Searching for protons with a kinetic energy of a few hundred
keV with relative intensity $10^{-8}$ is challenging in a radioactive
beam environment, so we instead detect the decay product,
$^{10}$Be with a halflife of $1.5 \cdot 10^6$ y, that exists only in minute
quantities on earth. To reach the needed sensitivity we must employ
state-of-the-art AMS.  It is also crucial to limit the amount of
contaminants in the samples so sample collection took place at
ISOLDE's high-resolution mass separator.  The resolution from the
magnetic separation stage is supplemented by the electrostatic beam
transport at ISOLDE similar to, but at lower resolution than, the
separation stages in AMS facilities.  A first attempt was made in
2001 and the results were published recently \cite{Bor13}. The signal
was not sufficiently strong to be clearly separated from background
and gave a $\beta$p branching ratio of $(2.5 \pm 2.5) \cdot 10^{-6}$,
significantly above the published theoretical expectations.  Due to
improvements both in production of $^{11}$Be and AMS detection of
$^{10}$Be, the current collection was performed in December 2012 and
resulted in three samples.

\subsection{Sample collection}
The $^{11}$Be activity was produced by bombarding a UC
target with 1.4 GeV protons. The products were ionized in a laser ion
source, which provided element selectivity, mass separated in the
ISOLDE high-resolution separator, and guided through several collimators
to the collection point where they were implanted at 60 keV in a small
copper plate (15$\times$20$\times$2 mm). A high-purity coaxial
Ge-detector placed 40 cm downstream behind a lead shielding monitored
the collection rate. The Ge-detector was energy and efficiency
calibrated with standard sources of $^{60}$Co, $^{152}$Eu and
$^{228}$Th.  The main lines in the $\gamma$ spectrum recorded during
$^{11}$Be collection are the 2124 keV line from the decay of $^{11}$Be
and the 511 keV line from positron annihilation.  The overall
efficiency at 2124 keV is found to be $(2.0 \pm 0.2)\cdot10^{-5}$. A
second line from the decay at 2895 keV was also used to check the
overall amount of $^{11}$Be. The two determinations gave about the
same precision, the one from the 2124 keV line being dominated by
systematic uncertainties in the efficiency and the one from the 2895
keV being dominated by statistical uncertainties, and were internally
consistent leading to a final value for the amount of collected
$^{11}$Be in the main sample (S1) of $(1.447 \pm 0.055) \cdot
10^{12}$. This includes a correction for dead time of 2.8\%,
determined from the ratio of accepted to total number of triggers.

As cross-checks two other samples were collected:
sample S2 at the mass position of
$^{11}$Li (0.02 mass units heavier than $^{11}$Be) where an upper
limit of $3 \cdot 10^6$ could be determined for the number of atoms collected
(corresponding to a $^{11}$Li yield below 625/s which is reasonable)
and, for one second
only, sample S3 at the $^{10}$Be mass position where an estimate of the
current of 3.5 pA (uncertain by a factor two) converts into $2.2 \cdot 10^7$ atoms.
According to SRIM calculations \cite{SRIM} about 6\% of all Be ions
implanted in Cu at 60 keV energy will backscatter out of the
sample. Most of the backscattered ions are expected to remain close to
the sample so $\gamma$-rays from their decays will be seen as well,
although the decay products are not retained in the sample. This gives
a correction which we estimate to be $4 \pm 4$\%.

\subsection{Accelerator mass spectrometry}
The $^{10}$Be accelerator mass spectrometry (AMS) measurements were
performed at the Vienna Environmental Research Accelerator (VERA) at
the University of Vienna. VERA is a dedicated AMS facility based on a
NEC 3 MV pelletron tandem accelerator. A new scheme for $^{10}$Be
using a passive foil absorber in front of a gas ionization chamber
detector was employed. In this way the detection efficiency for
$^{10}$Be atoms is increased significantly.

According to TRIM simulations \cite{SRIM} the maximum implantation
depth of $^{11}$Be in our copper plate catcher was below 1 $\mu$m. 
To reduce the amount of material to
be dissolved only the surface layer of each irradiated copper plate
was leached in nitric acid. A second leaching was performed 
to verify the blank level of the irradiated copper
plate. The second leaching of sample S3 did not produce enough BeO
for a measurement. For
samples S1 and S2 the values of the
second leachings were consistent with a blank sample. This shows that
the material was sitting in the surface, as expected for an implanted
sample, and not due to a bulk contamination. An amount of 
359 $\mu$g (uncertainty 3\%) $^9$Be carrier was added to the solution to reach a
$^{10}$Be/$^9$Be isotopic ratio in the range of 10$^{-16}$--10$^{-11}$. In the next step
the solution was treated with ammonium hydroxide to precipitate the
beryllium as beryllium hydroxide (Be(OH)$_2$). The dissolved copper
remains in the solution in this step. The beryllium hydroxide was
dried out by heating in an oven at 900$^{\circ}$C for at least 8 hours forming
beryllium oxide (BeO). The BeO powder was mixed 1:1 with high purity
copper powder and pressed into sample holders
and mounted together with standard and blank material in a MC-SNICS
type Cesium sputter ion source.
Blank is the pure phenakite material directly
pressed into a sample holder. A separate sample, S-blank, went through the chemistry
preparation to check for the amount of $^{10}$Be introduced during the
chemical sample preparation.
BeO$^-$ was extracted from the ion source and
stripped in the terminal of the tandem accelerator to Be$^{2+}$,
resulting in a total ion energy of 2.4 MeV. After
further mass separation by a sector magnetic analyzer and an
electrostatic analyzer the remaining particles are sent to a gas
ionization chamber detector with a two-split anode for particle
identification. A silicon nitride foil stack as a passive absorber was
installed in front of the detector. This foil stack
prevents the isobaric background $^{10}$B from entering into the
detector: The energy loss of boron in the foil stack is slightly
larger compared to beryllium. By selecting the right foil thickness
and carefully tuning the particle energy the boron ions are stopped in
the foil stack whereas the beryllium ions can enter the detector. 

\begin{table}
\centering
\caption{Results of the AMS measurement. S1 to S3 denote the
irradiated samples. 1st and 2nd correspond to the first or second
leaching. Blank and S-blank are control samples without activity.}
\label{tab:ams}       
\begin{tabular}{lcc}
 \hline
Sample  & $^{10}$Be/$^9$Be ratio  & $^{10}$Be atoms \\ \hline
S1-1st  & $(4.87 \pm 0.13) \cdot 10^{-13}$  & $(1.17 \pm 0.05) \cdot 10^7$ \\
S1-2nd & $(1.26 \pm 0.56) \cdot 10^{-15}$  & $(3.03 \pm 1.35) \cdot 10^4$ \\
S2-1st  & $(3.10 \pm 0.94) \cdot 10^{-15}$  & $(7.45 \pm 2.27) \cdot 10^4$ \\
S2-2nd & $(4.4\pm 3.1) \cdot 10^{-16}$ & $(1.06 \pm 0.75) \cdot 10^4$ \\
S3-1st  & $(1.54 \pm 0.03) \cdot 10^{-12}$ & $(3.70 \pm 0.13) \cdot 10^7$ \\
S-blank & $(4.9 \pm 3.4) \cdot 10^{-16}$ & $(1.18 \pm 0.82) \cdot 10^4$ \\
blank    & $(1.3 \pm 1.3) \cdot 10^{-16}$ & $(3.12 \pm 3.12) \cdot 10^3$ \\ \hline
\end{tabular}
\end{table}

The final results are given
in table \ref{tab:ams}. The amount of atoms in sample S3 agrees with
the estimation from the implantation current. The number for sample S2
is consistent with the lack of observed $\gamma$-rays from the decay
of $^{11}$Li. The number for sample S1, the $^{11}$Be sample, is
$(1.170 \pm 0.047) \cdot 10^7$.

\begin{figure}[thb]
\centering
     \includegraphics[width=7cm,clip]{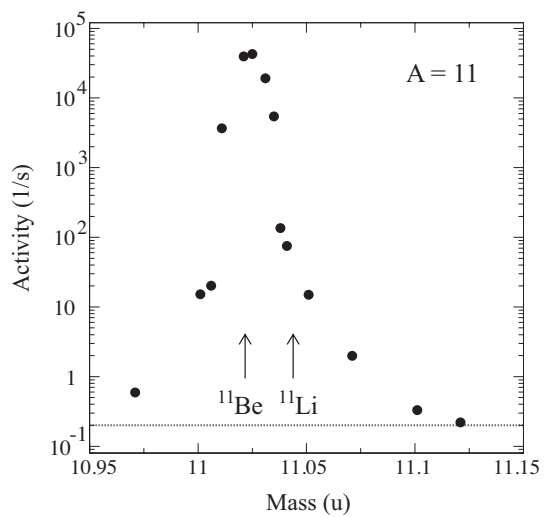} 
\caption{Mass scan of the ISOLDE high-resolution
  separator across the $^{11}$Be position. The beta activity measured
  is shown versus the mass with positions indicated for $^{11}$Be and the
  possible contaminant $^{11}$Li. The horizontal line marks the
  detection limit of 0.2/s.}
\label{fig:scan} 
\end{figure}

\subsection{Possible contaminants}
Contaminations in our sample might arise due to tails of the neighbouring
activities $^{10}$Be or $^{11}$Li, whose decay also
produces $^{10}$Be. Both possibilities are ruled out by the low
recorded number of atoms for the $^{11}$Li sample (S2). The ISOLDE mass
separator profile was found by measuring the beta activity as the mass
settings were changed around the nominal $^{11}$Be mass, see
figure~\ref{fig:scan}. The release function of this specific target
and ion source combination was measured first, which allows to combine
measurements with different collection times relative to proton impact
on target. In this way the sensitivity was increased and the activity
could be followed down to the $10^{-5}$ level that
occured at a mass difference of 0.05 mass units.
The only remaining way for $^{10}$Be to appear on the 
$^{11}$Be position is as the molecule $^{10}$Be$^1$H, but
this molecule is unlikely to be formed in the target and to survive
through the laser ion source since its ionization energy of 8.22 eV
\cite{Bub07} is much higher than its dissociation energy of 3.26 eV.
Nevertheless, we have re-checked the data from an earlier
experiment on $^{12}$Be \cite{Dig05} and were able to put limits on
the amount of $^{11}$Be$^1$H (from the $\beta\alpha$ branch) that would
correspond in our current case to a $^{10}$Be$^1$H intensity less than $2\cdot10^{-6}$
of $^{11}$Be. Our conditions should be
better, partly due to higher laser ionization power, partly due to the
beam passing through a gas-filled RFQ cooler, both
effects that would enhance molecular break-up. We therefore conclude
that we have observed the $^{11}$Be($\beta$p) decay via detection of
the final nucleus $^{10}$Be. The observed intensity converts to a branching
ratio of $(8.4 \pm 0.6) \cdot 10^{-6}$.

\section{Discussion}

The experimentally found branching ratio is surprisingly large, but
consistent with the outcome of the first experiment. 
If the strength in $^{11}$Be($\beta$p) was as
broadly distributed as in $^{11}$Li($\beta$d) we would expect the $B_{GT}$ within
the Q-window to be less than 0.1, which would not be sufficient to
explain the decay rate. We therefore turn to a simple model for the
decay along the lines of the direct decay calculations in \cite{Bay11,Zhu95},
details of the calculations are reported elsewhere \cite{Rii14}.  

\begin{figure}[thb]
\centering
    \includegraphics[width=7.cm,clip]{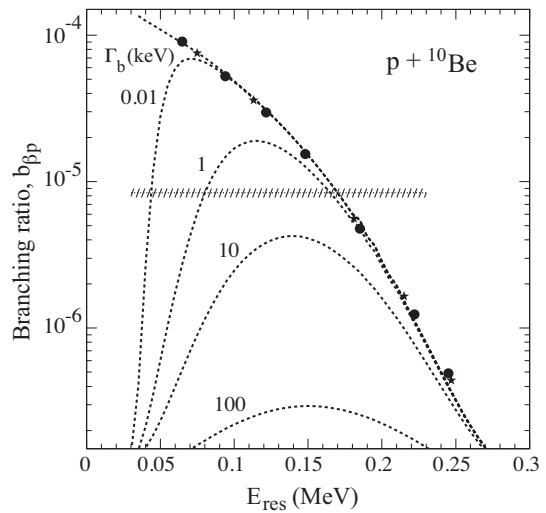} 
\caption{The calculated branching ratio for decay into p+$^{10}$Be is
   shown as a function of the centre-of-mass energy of the resonance.
   The stars mark results of calculations with square well
   potentials, the filled circles are results from Woods-Saxon potentials.
   The curves arise from integrating the R-matrix expression in
   equation (\protect\ref{eq:Rmatrix}) for different widths for other
   decay channels.
  The horizontal band indicates the experimentally found branching ratio.}
\label{fig:th_br} 
\end{figure}

The basic assumption is that the beta decay proceeds as an essentially
detached decay of the halo neutron into a proton. The initial and
final state wavefunctions are calculated as single-particle states in
square-well or Woods-Saxon potentials with the final state spectrum discretized
by imposing a large confining radius at 1000 fm.
The overlap of the wavefunctions gives the beta strength $B_{GT}$ and the
decay rate is found from equation (\ref{eq:ft}). The final total branching
ratio for beta-delayed proton emission depends strongly on the
strength of the potential between the final state proton and
$^{10}$Be. For most potential strengths the branching ratio will
indeed be a few times $10^{-8}$, as in other calculations, but in a
limited range the beta strength will be concentrated within the
Q-window. Effectively, in this range the proton formed in the decay
interacts strongly with the remaining $^{10}$Be and forms a
resonance-like structure; as a consequence it emerges with a quite
well defined energy. The branching ratios obtained for this set of
parameters are shown in figure \ref{fig:th_br} as a function of the
energy of the resonance.

The simple model neglects isospin. The lowest $T=3/2$ states are
situated slightly more than 1 MeV above the $Q_{\beta p}$ window.
They are members of isospin multiplets that
include the $^{11}$Be ground state and first excited state neutron
halos. The data indicate \cite{Jon98} that the intermediate states
($|T_z|$ of 1/2) in these multiplets have good
total isospin rather than a composition with just one proton (or
neutron) plus core. We therefore expect that realistic
final state wave functions in our case, with
$T=1/2$, also should have good isospin. Standard isospin
coupling then predicts that the state should be proton plus $^{10}$Be
with weight 2/3 and neutron plus $^{10}$B(T=1) with weight 1/3.
Our calculated decay probabilities must therefore be corrected by
a factor 2/3.
A further reduction factor about 0.7 is due to the
initial $^{11}$Be wavefunction containing several configurations
\cite{Sch12}. The overall scaling factor on the theory, included in
figure \ref{fig:th_br}, is therefore about 0.5.

Could this be an established resonance in $^{11}$B ? The known states
\cite{TUNL11} in this region mainly couple to the $\alpha$-particle
channel (with partial widths around 100 keV) and only one, a state at
$11450 \pm 17$ keV, may have spin-parity that allows emission of an
s-wave proton --- the others will have angular momentum barriers that
will suppress proton emission.  Decays through levels that have other
sizeable decay channels ($\alpha$ emission or, for very narrow levels,
$\gamma$ emission; in principle triton emission could also occur)
would therefore only contribute to the proton channel with probability
$\Gamma_p/\Gamma_{tot}$. Since $\Gamma_{\gamma}$ for the $1/2^+$,
$T=3/2$ state at 12.55 MeV (that apart from isospin should be similar
in structure to our state) is about 10 eV \cite{TUNL11}, and even a
small admixture into our state of other $1/2^+$ levels is likely to
give a $\Gamma_{\alpha}$ at least of the same magnitude, we shall
assume the width for other decay channels $\Gamma_b =
\Gamma_{\gamma} + \Gamma_{\alpha}$ to be larger than 0.01 keV.

To take these effects into account calculations were also made within
the R-matrix approach \cite{Bar88}, but in
a simplified version where e.g.\ other decay channels are
approximated as having a constant width $\Gamma_{\alpha}$ over the
energy window, see \cite{Rii14} for details.  Converting the decay
rate into a differential branching ratio gives the following
expression:
\begin{equation}  \label{eq:Rmatrix}
     \frac{\mathrm{d}b}{\mathrm{d}E} = t_{1/2} \frac{g_A^2}{K} 
         \frac{B_{GT}\Gamma_p/2\pi}{(E_{res}-E)^2+\Gamma_{tot}^2/4} f(Q-E)\,,
\end{equation}
where $\Gamma_{tot} = \Gamma_b+\Gamma_p$, $\Gamma_p =
2P\gamma^2$, $P$ is the standard (energy-dependent) penetrability
factor and $\gamma^2$ the maximal reduced width.  Integration over
the Q-window gives the total branching ratios shown in figure
\ref{fig:th_br} as a function of resonance position $E_{res}$ for
different values of $\Gamma_b$. The branching ratios agree well
with the ones from the simple model. It is clear that all
known levels are too wide to fit and that a $\Gamma_b$ above 0.01 keV
gives a lower limit on the $B_{GT}$ of about 0.3 with an upper limit
given by the theoretical maximum of 3.

\section{Conclusion}

We have observed beta-delayed proton emission for
the first time in a neutron-rich nucleus. The unexpectedly high
decay rate can only be understood within current
theory if the decay proceeds through a new single-particle resonance
in $^{11}$B that is strongly fed in beta-decay. 
The $B_{GT}$-value could be as high as that of a free neutron decay.
A natural
interpretation would be peripheral beta decay of the halo neutron in
$^{11}$Be into a single-proton state. This appears to be a simpler
process than the $\beta$d decays of the two-neutron halo nuclei $^6$He
and $^{11}$Li. Although the halo structure must be important for the $\beta$p
decay mode, the large value of $B_{GT}$ may be related to large values
found in other (non-halo) near-dripline nuclei \cite{Bor91} and point
to a more widespread change of beta-decay patterns at least in light
nuclei in line with some predictions \cite{Sag93}.

\section*{Acknowledgements}
We thank the ISOLDE group for
the successful operation of the HRS separator at very high
resolution and Aksel Jensen for discussions on the theoretical
interpretation. We acknowledge support from the European Union
Seventh Framework through ENSAR (contract no. 262010), from
Austrian Science Fund (FWF) P22164-N20, from
Spanish MINECO through projects FPA2010-17142 and FPA2012-32443, and 
CPAN Consolider CSD-2007-00042.

\end{document}